\documentclass[RNAAS]{aastex62}
\usepackage{hyperref}
\usepackage{newunicodechar,graphicx}

\DeclareRobustCommand{\okina}{%
  \raisebox{\dimexpr\fontcharht\font`A-\height}{%
    \scalebox{0.8}{`}%
  }%
}
\newunicodechar{ʻ}{\okina}

\newcommand{\unsw}{School of Physics, University of New South Wales, Sydney, NSW 2052, Australia}

\begin{document}

\title{Systematics-insensitive Periodogram for finding periods in TESS observations of long-period rotators}

\correspondingauthor{Christina Hedges}
\email{christina.l.hedges@nasa.gov}

\author[0000-0002-3385-8391]{Christina Hedges}
\affiliation{Bay Area Environmental Research Institute, P.O. Box 25, Moffett Field, CA 94035, USA}
\affiliation{NASA Ames Research Center, Moffett Field, CA}

\author[0000-0003-4540-5661]{Ruth Angus}
\affiliation{Department of Astrophysics,
American Museum of Natural History, 200 Central Park West, New York, NY 10024}
\affiliation{Center for Computational Astrophysics, Flatiron Institute, 162 5th Avenue, New York, NY, 10010}

\author[0000-0002-3306-3484]{Geert Barentsen}
\affiliation{Bay Area Environmental Research Institute, P.O. Box 25, Moffett Field, CA 94035, USA}
\affiliation{NASA Ames Research Center, Moffett Field, CA}

\author[0000-0003-2657-3889]{Nicholas Saunders}
\affiliation{Institute for Astronomy, University of Hawaiʻi at M\=anoa, 2680 Woodlawn Drive, Honolulu, HI 96822, USA}

\author[0000-0001-7516-8308]{Benjamin~T.~Montet}
\affiliation{\unsw}

\author[0000-0002-4020-3457]{Michael Gully-Santiago}
\affiliation{The University of Texas at Austin Department of Astronomy, 2515 Speedway, Stop C1400, Austin, TX 78712, USA}



\section{}
NASA's TESS mission \citep{tess} has produced high precision photometry of millions of stars to the community. The majority of TESS observations have a duration of $\approx$27 days, corresponding to a single observation during a TESS sector. A small subset of TESS targets are observed for multiple sectors, with approximately 1-2\% of targets falling in the Continuous Viewing Zone (CVZ) during the prime mission \citep{yield}, where targets are observed continuously for a year. These targets are highly valuable for extracting long period rotation rates, which can be linked to stellar ages.

The TESS spacecraft orbits the earth every $\approx$14 days, after which there is a short data-downlink. Typical observations last 11-13 days between downlinks/pointing changes. TESS experiences significant scattered light during each orbit. These factors cause a systematic signal in the TESS data, which can make it difficult to 1) ``stitch'' time-series data together across data gaps between sectors and after downlinks 2) identify long-period trends in the data, particularly at periods of $\gtrsim$ 28 days. We present a method to create a Lomb-Scargle periodogram, while simultaneously detrending these TESS systematics. This method is similar to the work presented in \cite{sip} for detrending systematics in the NASA Kepler/K2 dataset, and we direct readers to that work for a full discussion. Similarly to \cite{sip}, we refer to this periodogram as a Systematics-insensitive Periodogram (SIP).

Our implementation of SIP is a simple linear model, consisting of regressors to remove instrument systematics, and a sinusoid component to fit a power spectrum as a function of period (i.e. a periodogram). The principles behind our linear model are presented simply and efficiently in \cite{themagic}. We build our systematic regressors using the same principles as the K2 \texttt{EVEREST} pipeline \citep{everest}. Our method uses the following procedure: 1) using the TESS Target Pixel File data, and apertures assigned by the TESS Pipeline \citep{spoc, spoc2}, we build Simple Aperture Photometry (SAP) light curves of the target, including all scattered light contributions (i.e. not using the pipeline provided background correction). 2) we build an estimate of the background using the first 3 principal components of the pixel time series outside of the aperture. This creates a 3 by $t$ matrix, where $t$ is the number of time points in the dataset. We then include a column of ones to account for mean offsets, making a 4 by $t$ matrix. These are the systematics regressors. 3) We duplicate the regressors for every 14 day time-series segment between data-downlinks. For each segment, we then set the values in the regressors to zero at all \textbf{other} segments. This creates a sparse matrix with size $4s$ by $t$, where s is the number of segments. This matrix has block-diagonal structure, and has values only during each 14 day segment. 4) We create a simple sine and cosine curve at a given period, evaluated at all time points, using \texttt{astropy}'s \texttt{LombScargle} module. 5) Using \texttt{lightkurve}'s \texttt{RegressionCorrector} framework, we fit the systematics regressors and sinusoid components simultaneously to the SAP time-series flux data, including regularization terms to avoid overfitting. This process is run for every period of interest. The ``power" in the periodogram is defined as the amplitude of the sinusoid at each period. TESS data from the CVZ can contain 100,000+ data points, and calculating thie SIP can become expensive in memory. To avoid this, we rely on \texttt{lightkurve}'s \texttt{SparseDesignMatrix} class to keep memory usage low \citep{lk}. 

An example of the results of our SIP method is shown in Figure 1. The tool used to generate this SIP is available online\footnote{\href{https://doi.org/10.5281/zenodo.4300754}{doi.org/10.5281/zenodo.4300754}}\footnote{\href{https://github.com/christinahedges/TESS-SIP}{github.com/christinahedges/TESS-SIP}} as a pip installable Python tool. The method presented in this note can be easily modified to stitch together TESS CVZ observations using 1) different systematics models (e.g. models for the telescope jitter), or 2) different models for stellar variability (e.g. a simple basis-spline model for non-periodic variability).

\begin{figure}[h!]
\begin{center}
\includegraphics[scale=0.85,angle=0]{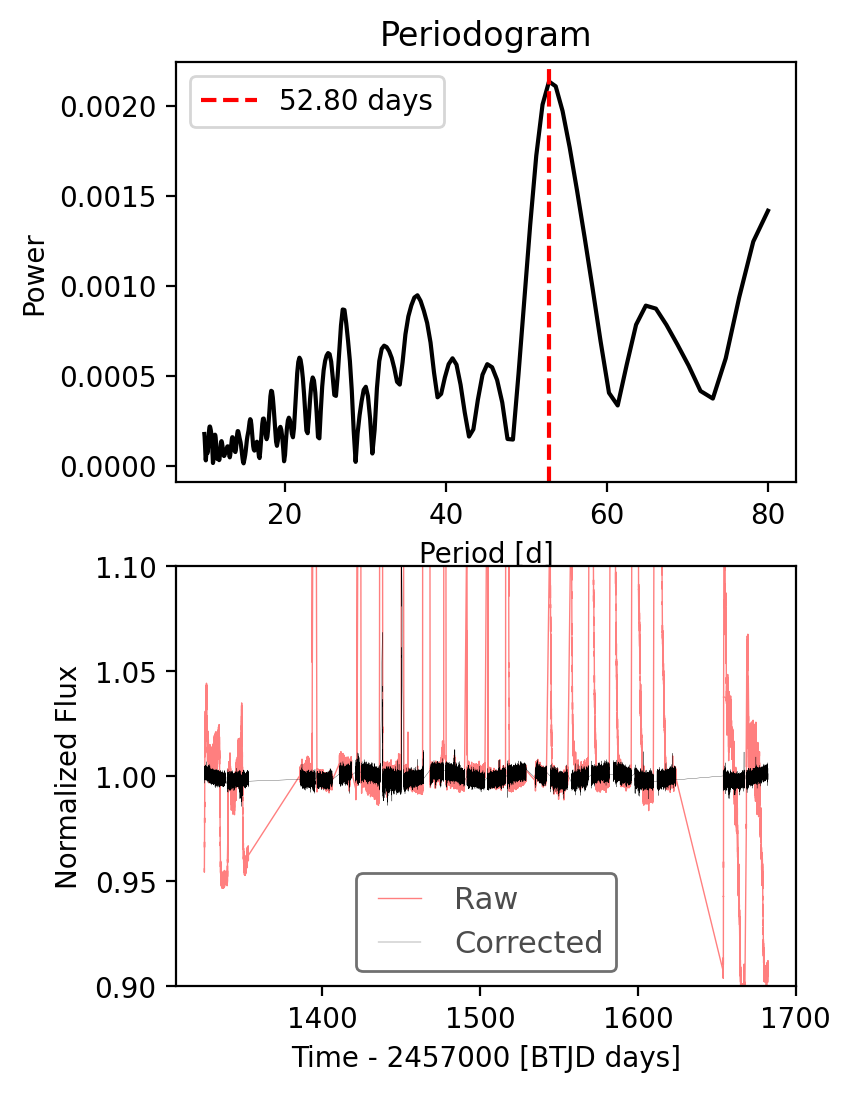}
\caption{Example of the Systematics-insensitive Periodogram generated for target TIC 150428135 (TOI-700). Black shows the detrended light curve, with the best fit systematics at the maximum power period removed. This target has a significant, long period rotation, at 52 days, longer than the orbital period of the TESS spacecraft (28 days). Without using these strategies to mitigate instrument systematics, it is difficult to recover rotation rates at periods close to/greater than the TESS orbital period.}
\end{center}
\end{figure}


\acknowledgments
The SIP project was developed in part at the ``online.tess.science'' meeting, which took place globally in 2020 September. This research made use of Lightkurve, a Python package for Kepler and TESS data analysis (Lightkurve Collaboration, 2018). This research made use of Astropy,\footnote{\href{http://www.astropy.org}{http://www.astropy.org}} a community-developed core Python package for Astronomy \citep{astropy:2013, astropy:2018}.

\bibliography{bib}{}
\bibliographystyle{aasjournal}

\end{document}